\documentclass[prb,twocolumn]{revtex4}
\usepackage{amsmath}
\usepackage{graphicx}

\begin{document}

\title{Percolation transition and dissipation in quantum Ising magnets}

\author{Jos\'{e} A. Hoyos}
\author{Thomas Vojta}

\affiliation{Department of Physics, University of Missouri-Rolla, Rolla,
Missouri 65409, USA.}

\begin{abstract}
We study the effects of dissipation on a randomly diluted transverse-field
Ising magnet close to the percolation threshold. For weak transverse fields,
a novel percolation quantum phase transition separates a superparamagnetic
cluster phase from an inhomogeneously ordered ferromagnetic phase. The
properties of this transition are dominated by large frozen and slowly
fluctuating percolation clusters. This leads to a discontinuous
magnetization-field curve and exotic hysteresis phenomena as well as highly
singular behavior of magnetic susceptibility and specific heat. We compare our
results to the smeared transition in generic dissipative random quantum Ising
magnets. We also discuss the relation to metallic quantum magnets and other
experimental realizations.
\end{abstract}


\maketitle



In diluted quantum magnets, the combination of geometric and quantum fluctuations can
lead to unconventional low-temperature properties such as singular thermodynamic
quantities in quantum Griffiths phases as well as exotic scaling at the quantum phase
transitions (QPTs) between magnetic and nonmagnetic ground states
\cite{Fisher92,Fisher95,ThillHuse95,YoungRieger96} (for a recent review see, e.g., Ref.\
\onlinecite{Vojta06}). In many real systems, the magnetic degrees of freedom are coupled
to an environment of ``heat bath'' modes.  The dissipation caused by the bath is known to
qualitatively change the properties even of a single quantum
spin.\cite{LCDFGZ87,Weiss_book93,Hewson_book93,VojtaM06}.

\begin{figure}
\includegraphics[width=0.85\columnwidth]{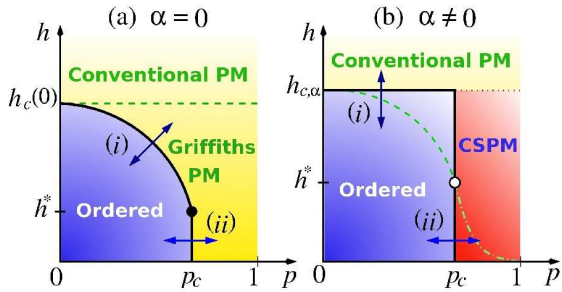}
\caption{Schematic ground state phase diagrams of diluted quantum Ising magnets without
(a) and with (b) dissipation. CSPM is the classical superparamagnetic phase. The dashed
line in (b) marks the crossover between homogeneous and inhomogeneous order in the
smeared transition scenario.
\label{cap:Phase-diagram}}
\end{figure}

In this Letter, we show that dissipation dramatically changes phases and phase
transitions of a randomly diluted quantum Ising magnet, as is illustrated by Fig.\
\ref{cap:Phase-diagram}.
The behavior without dissipation is well
understood:\cite{Harris74b,Stinchcombe81,Santos82,SenthilSachdev96} Magnetic long-range
order exists for dilutions $p$  up to the lattice percolation threshold $p_c$ as long as
the transverse field $h_x$ is below a critical field $h_c(p)$. For $p>p_c$, long-range
order is impossible because the system is decomposed into noninteracting finite-size
clusters. There are two QPTs, separated by a multicritical point at $(p_c, h^\ast)$. The
transition across $h_c(p)$ for $0<p<p_c$ (transition (i) in Fig.\
\ref{cap:Phase-diagram}a) falls into the generic random quantum Ising universality class
and is characterized by an infinite-randomness fixed point.\cite{Fisher92,Fisher95} The
transition across $p_c$ at $h_x<h^\ast$ (transition (ii) in Fig.\
\ref{cap:Phase-diagram}a) is an unusual kind of percolation QPT. \cite{SenthilSachdev96}
The paramagnetic phase consists of two regions: a conventional gapped region for
$h_x>h_c(0)$ and a gapless quantum Griffiths region for $h_x<h_c(0)$ where rare regions
lead to power-law thermodynamic singularities.

In the presence of weak ohmic dissipation, the phase diagram changes qualitatively since
the quantum dynamics of sufficiently large clusters
freezes.\cite{MillisMorrSchmalian01,MillisMorrSchmalian02,CastroNetoJones00} This leads
to the appearance of an unusual classical superparamagnetic cluster phase. For $p>p_c$,
it is stabilized down to zero temperature because different clusters are strictly
decoupled. In contrast, for $p<p_c$ weakly coupled frozen rare regions align. Magnetic
long-range order thus extends to the clean critical field $h_{c,\alpha}$, and the
field-driven phase transition (i) is smeared.\cite{Vojta03a,Vojta03b,SchehrRieger06}

In this Letter, we focus on the novel percolation QPT at small fields $h_x$ (transition
(ii) in Fig.\ \ref{cap:Phase-diagram}b). Our results can be summarized as follows: The
total magnetization consists of a coherent part $m_\infty$ and an incoherent part $m_{\rm
st}$. The critical behavior of the coherent part is given by classical percolation theory
$m_\infty \sim |p-p_c|^{\beta_c}$ with $\beta_c$ the classical percolation exponent.
Somewhat unexpectedly, $m_\infty + m_{\rm st}$ (which is measured in an infinitesimal
ordering field) is \emph{noncritical} at $p_c$. The interplay between the two parts leads
to unusual magnetization-field hysteresis. At low temperatures $T$, the susceptibility
has a Curie contribution from the frozen clusters that diverges as $\chi \sim
|p-p_c|^{-\gamma_c}/T$ with the classical percolation exponent $\gamma_c$. Fluctuating
clusters provide a subleading $1/(T\ln^2T)$ contribution. In contrast, the
low-temperature specific heat is dominated by fluctuating clusters, resulting in a
logarithmic temperature dependence, $C \sim \ln^{-2}(1/T)$. The logarithmic terms in
$\chi$ and $C$ exist on both sides of $p_c$ and their prefactors are noncritical at this
point .


Our starting point is a $d$-dimensional ($d\ge 2$) site-diluted transverse-field Ising
model\cite{Harris74b,Stinchcombe81,Santos82,SenthilSachdev96}
\begin{equation}
H_I= -J\sum_{\langle i,j\rangle
}\epsilon_{i}\epsilon_{j}\sigma_{i}^{z}\sigma_{j}^{z}-h_x \sum_{i}\epsilon_{i}\sigma_{i}^{x}~,
\label{eq:H}
\end{equation}
a prototypical disordered quantum magnet. The Pauli matrices $\sigma_{i}^{z}$ and
$\sigma_{i}^{x}$ represent the spin components at site $i$, the exchange interaction $J$
couples nearest neighbor sites, and the transverse field $h_x$ controls the quantum
fluctuations. Dilution is introduced via random variables $\epsilon_{i}$ which can take
the values 0 and 1 with probabilities $p$ and $1-p$, respectively. We now couple each
spin to a local ohmic bath of harmonic
oscillators,\cite{CugliandoloLozanoLozza05,SchehrRieger06}
\begin{equation}
H = H_I + \sum_{i,n}  \left[\nu_{i,n}a_{i,n}^{\dagger}a_{i,n}+ \frac 1 2
\lambda_{i,n}\sigma_{i}^{z} (a_{i,n}^{\dagger}+a_{i,n}) \right],\label{eq:Hamiltonian}
\end{equation}
where $a_{i,n}$ and $a_{i,n}^{\dagger}$ are the annihilation and creation operators of
the $n$-th oscillator coupled to spin $i$; $\nu_{i,n}$ is its frequency, and
$\lambda_{i,n}$ is the coupling constant. All baths have the same spectral function
${\cal E}(\omega)=\pi \sum_{n}\lambda_{i,n}^{2} \delta (\omega-\nu_{i,n})/\nu_{i,n}=2\pi
\alpha\omega e^{-\omega/\omega_{c}}$ with $\alpha$ the dimensionless dissipation strength
and $\omega_{c}$ the cutoff energy. Such local ohmic dissipation can be realized in good
approximation in magnetic nanoparticles in an insulating host.\cite{Wernsdorfer01} It
also applies to resistively shunted SQUIDs\cite{LCDFGZ87} and qualitatively (possibly
with a different spectral density ${\cal E}(\omega)$) to molecular magnets weakly coupled
to nuclear spins\cite{ProkofevStamp00} (see, e.g., the spin-1/2 molecular complex
V$_{15}$, Ref.\ \onlinecite{CWMBB00}).


Let us start by considering a single percolation cluster of $s$ occupied sites. Without
dissipation, and for small transverse fields, $h_x \ll h_{c,\alpha} \sim J$, its lowest
two energy levels correspond to the states of a single effective Ising spin with a moment
proportional to $s$ in an effective transverse field (tunneling matrix element)
$\Delta_s$. For large $s$, $\Delta_s$ can be estimated in $s$-th order of perturbation
theory to be $h_x e^{-Bs}$, with $B$ a constant of order
$\ln(J/h_x)$.\cite{SenthilSachdev96} All other levels are separated by energies of order
$J$ and can thus be neglected for the low-energy physics. This means the cluster tunnels
coherently between \emph{up} and \emph{down} with a tunneling frequency $\Delta_s$.

The effects of the heat bath on the cluster can be worked out in 2nd order perturbation
theory in $h_x$. We find that the effective spin of the cluster feels a single ohmic bath
with dissipation strength $A s \alpha$ because the number of oscillators coupling to the
cluster is proportional to $s$.\cite{HoyosVojta_unpublished}  Here, $A$ is an
$h_x$-dependent constant which we suppress as it can be absorbed in the cluster moment.
Analogous results have been obtained for the dissipative random quantum Ising
chain\cite{SchehrRieger06} and itinerant Ising magnets.\cite{MillisMorrSchmalian02}
Each percolation cluster thus behaves like an ohmic spin-boson
model:\cite{LCDFGZ87,Weiss_book93} For strong dissipation, $s \alpha>\alpha_c \approx 1$
\footnote{The value $\alpha_c=1$ is the result for $\Delta_s/\omega_c \to 0$. For finite
$\Delta_s$, $\alpha_c$ will be larger than one. We will neglect this difference as it is
of no importance for our calculations.}, the cluster is in the localized phase, i.e., the
renormalized tunneling matrix element $\Delta_{R,s}$ vanishes, and the magnetization is
frozen in one of the directions. Clusters with $s \alpha< 1$ are in the delocalized
phase, i.e., they still tunnel, but with a greatly reduced frequency. For $s \alpha\ll 1$
it can be estimated by adiabatic renormalization\cite{LCDFGZ87},
\begin{equation}
\Delta_{R,s} \sim \Delta_s \left({\Delta_s}/{\omega_{c}}\right)^{\alpha_s/(1-\alpha_s)} =
h_x\exp\left[ -{bs}/({1- s \alpha})\right]~, \label{eq:gap_renorm}
\end{equation}
with the constant $b=B+\alpha\ln(\omega_{c}/h_x)$. The QPT at $s \alpha=1$ is of
Kosterlitz-Thouless\cite{KosterlitzThouless73} type, analogous to the transition in the
classical $1/r^2$ Ising chain.\cite{Thouless69,Cardy81} Here, $\Delta_{R,s}$ plays the
role of the classical correlation length. Therefore, the functional form
(\ref{eq:gap_renorm}) remains valid as long as  $1-s \alpha > \Delta_s/\omega_c$.
However, very close to the transition, $1-s \alpha < \Delta_s/\omega_c$, eq.\
(\ref{eq:gap_renorm}) has to be replaced by\cite{Kosterlitz74}
\begin{equation}
\Delta_{R,s} \sim  h_x\exp\left( -{b^\prime s}/{\sqrt{1- s \alpha}}\right)~.
\label{eq:gap_KT}
\end{equation}


These results allows us to determine the phase diagram, Fig.\ \ref{cap:Phase-diagram}b,
for fixed small dissipation strength $\alpha$. For $h_x > h_{c,\alpha}$, our system is a
conventional gapped quantum paramagnet for all $p$ because not even the clean bulk system
orders. For $h_x < h_{c,\alpha}$ and $p>p_c$, the system is decomposed into finite-size
percolation clusters. The largest of these clusters are frozen and behave like classical
Ising magnets while the smaller ones fluctuate. The system is thus in a classical
superparamagnetic cluster phase. For $p<p_c$, ferromagnetic long-range order exists on
the infinite percolation cluster. The ferromagnetic phase extends up to the clean
critical field $h_{c,\alpha}$ for all $p<p_c$. This is a manifestation of the smeared
transition scenario:\cite{Vojta03a} The infinite percolation cluster contains, with small
probability, large impurity-free regions that freeze close to $h_{c,\alpha}$. Due to
their coupling to the rest of the infinite cluster they induce a highly inhomogeneous but
coherent magnetization already for transverse fields close to $h_{c,\alpha}$. For smaller
transverse fields there will be a crossover to homogeneous order on the infinite cluster
(see the dashed line in Fig.\ \ref{cap:Phase-diagram}b).


To study the percolation QPT, we need the distribution of cluster sizes as a function of
dilution $p$. According to percolation theory,\cite{StaufferAharony_book91} the number
$n_{s}$ of occupied clusters of size $s$ per lattice site obeys the scaling form
\begin{equation}
n_{s}\left( t\right) =s^{-\tau_c }f\left( ts^{\sigma_c }\right) .
\label{eq:percscaling}
\end{equation}
Here $t=p-p_c$, and $\tau_c $ and $\sigma_c$ are classical percolation exponents
\footnote{Classical percolation exponents carry a subscript $c$.}. The scaling function
$f(x)$ is analytic for small $x$ and has a single maximum at some $x_{\rm max}>0$. For
large $|x|$, it drops off rapidly
\begin{equation}
f(x) \sim \left\{ \begin{array}{lc} \exp\left(-(c_1 x)^{1/\sigma_c}\right) &
~(\textrm{for} ~x>0),\\ \exp\left(-(c_2 x)^{(1-1/d)/\sigma_c}\right) & ~(\textrm{for} ~
x<0),\end{array} \right. \label{eq:scaling_function}
\end{equation}
where $c_1$ and $c_2$ are constants of order unity. In addition, for $p<p_c$, there is an
infinite cluster containing a fraction $P_\infty \sim |p-p_c|^{\beta_c}$ of all sites.
All classical percolation exponents are
determined by $\tau_c$ and $\sigma_c$ including the correlation length exponent $\nu_c =({\tau_c -1})/{%
(d\sigma_c )}$, the order parameter exponent $\beta_c=(\tau_c-2)/\sigma_c$, and the
susceptibility exponent $\gamma_c=(3-\tau_c)/\sigma_c$.


The low-energy density of states (DOS) $\rho_{\rm dy}(\epsilon)$ of the dynamic clusters
($s \alpha <1$) can be determined by combining the single-cluster results
(\ref{eq:gap_renorm}) and (\ref{eq:gap_KT}) with the cluster size distribution
(\ref{eq:percscaling}) via $\rho_{\rm dy}(\epsilon) = \sum_{s<s_c}  n_s \,\delta
(\epsilon-\Delta_{R,s})$. Inserting (\ref{eq:gap_renorm}) for $\Delta_{R,s}$ yields
\begin{eqnarray}
\rho_{\rm dy}(\epsilon) =  n_{s(\epsilon)} \, b / \left[ \epsilon \,( b+\alpha
\ln(h_x/\epsilon))^2\right]~, \label{eq:DOS}
\end{eqnarray}
where $n_{s(\epsilon)}$ is the density of clusters of size
$s(\epsilon)=\ln(h_x/\epsilon)/[b+\alpha\ln(h_x/\epsilon)]$. Eq.\ (\ref{eq:DOS}) contains
the full crossover between quantum Griffiths behavior at higher energies and
damping-dominated behavior at low energies. For $\epsilon>\epsilon_{\rm cross} =h_x
e^{-b/\alpha}$, the DOS simplifies to a power-law form, $\rho_{\rm dy}(\epsilon) \sim
\epsilon^{c/b-1}$, characteristic of quantum Griffiths behavior.\cite{ThillHuse95} In
contrast, for $\epsilon<\epsilon_{\rm cross}$, the energy dependence is even more
singular,
\begin{equation}
\rho_{\rm dy}(\epsilon) \sim {n_{s_c}} /[{\epsilon \ln^\phi(h_x/\epsilon)}]~,
\label{eq:DOS_asymp}
\end{equation}
with $\phi=2$. At very low energies, in the asymptotic region of the Kosterlitz-Thouless
transition, $\Delta_{R,s}$ is given by (\ref{eq:gap_KT}) which leads to $\phi=3$. The
functional form (\ref{eq:DOS_asymp}) of the low-energy DOS is the same on both sides of
the percolation threshold since it is caused by clusters of \emph{finite size} $s \approx
s_c=1/\alpha$. Moreover, the prefactor is noncritical at $p_c$ because the cluster size
distribution $n_s$ is an analytic function of $p$ for any finite $s$.

\begin{figure}
\begin{center}
\includegraphics[width=0.74\columnwidth]{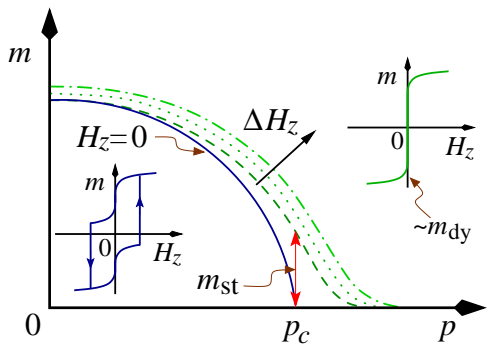}
\end{center}
\caption{Zero-temperature magnetization as a function of dilution $p$ for different
ordering fields $H_z$. The solid line is the coherent magnetization at $H_z=0$ (infinite
cluster only). The dashed line is for an infinitesimal field. The insets show the
singular field dependence of $m$ below and above $p_c$.} \label{cap:Magnetization}
\end{figure}

We now discuss the physics at the percolation transition in small transverse fields ($h_x
\ll h_{c,\alpha}$), starting with the total magnetization $m$. We have to distinguish the
contributions $m_{dy}$ from dynamic clusters, $m_{st}$ from frozen finite-size clusters,
and $m_{\infty}$ from the infinite percolation cluster, if any. For zero ordering field
$H_z$ in $z$-direction, $m_{dy}$ vanishes, because the dynamic clusters are in symmetric
superpositions of the \emph{up} and \emph{down} states. The frozen finite-size clusters
individually have a nonzero magnetization, but it sums up to $m_{st}=0$ because they do
not align coherently for $H_z=0$. The only coherent contribution to the total
magnetization is $m_{\infty}$. Since the infinite cluster is long-range ordered for small
transverse field $h_x$, its magnetization is proportional to the number $P_\infty$ of
sites in the infinite cluster, giving
\begin{equation}
m =m_\infty \sim P_\infty(p) \sim \left \{ \begin{array}{cc} |p-p_c|^{\beta_c} &
~(\textrm{for}~p<p_c),
\\ 0 & ~(\textrm{for}~p>p_c), \end{array} \right.  \label{eq:m}
\end{equation}
(solid line in Fig.\ \ref{cap:Magnetization}). The magnetization critical exponent is
given by its classical lattice percolation value $\beta_c$

The zero-temperature response of the magnetization to a small ordering field $H_z$ is
highly singular. Frozen finite-size clusters align parallel to the field already for
infinitesimal $H_z$, leading to a jump in $m(H_z)$ at $H_z=0$. The magnitude of the jump
is given by $m_{\rm st} \sim \sum_{s>s_c} n_s$. At the percolation threshold, $m_{\rm
st}$ is approximately given by $(1-p_c)s_c^{2-\tau_c}$, and it is exponentially small for
both $p \to 0$ and $p \to 1$. Somewhat surprisingly, the total magnetization in an
infinitesimal field, given by $m_\infty+m_{\rm st}$, is analytic (i.e., noncritical) at
$p=p_c$ (see dashed line in Fig.\ \ref{cap:Magnetization}). This follows from the fact
that only clusters with sizes below $s_c$ are \emph{not} polarized. Since the cluster
size distribution is analytic for any finite $s$, their total size is analytic at $p_c$,
and consequently $m_\infty+m_{\rm st}$ must be analytic, too.
To estimate the contribution $m_{\rm dy}$ of the dynamic clusters to the magnetization,
we integrate the magnetization of a single cluster, $H_z s^2(\epsilon)/(H_z^2
s^2(\epsilon) + \epsilon^2)^{1/2}$ over the DOS given in (\ref{eq:DOS}). For small fields
$H_z < \epsilon_{\rm cross}/s_c$, the main contribution comes from clusters with lowest
energies and sizes close to $s_c$, resulting in a highly singular dependence of $m_{\rm
dy}$ on the ordering field, $m_{\rm dy} \sim n_{s_c}/ \ln^{\phi-1}(h_x/H_z)$ for $H_z \to
0$. For higher fields, we find a crossover to quantum Griffiths behavior of the form
$m_{\rm dy} \sim H_z^{c/b}$.

As a consequence, our system displays unusual hysteresis (insets of Fig.\
\ref{cap:Magnetization}). For $p<p_c$, the long-range ordered infinite cluster
contributes a conventional single-domain hysteresis loop. Finite-size frozen clusters do
not show hysteresis, for arbitrarily small $H_z$, they relax with a nonzero
rate.\cite{LCDFGZ87} Thus, they contribute jumps in $m$ at $H_z=0$. The dynamic clusters
add logarithmic singularities at $H_z=0$. For $p>p_c$, there is no infinite cluster, but
the jump at $H_z=0$ as well as the logarithmic singularity survive. The above picture is
valid for adiabatically slow sweeps of the ordering field $H_z$. If $H_z$ is changed at a
finite rate, the largest clusters will fall out of equilibrium which leads to very
interesting time-dependent hysteresis phenomena that will be reported
elsewhere.\cite{HoyosVojta_unpublished}


The low-temperature susceptibility is dominated by the contribution $\chi_{\rm st}$ of
the static clusters, each of which has a Curie susceptibility of the form $s^2/T$.
Summing over all static clusters gives
\begin{equation}
\chi_{\rm st} \sim  \sum_{s>s_c} n_s  s^2/T \sim  |p-p_c|^{-\gamma_c}/T
\end{equation}
close to the percolation threshold. For $p\to 0$ and $p \to 1$, the prefactor of the
Curie term vanishes exponentially.
To determine the subleading contribution $\chi_{\rm dy}$ of the dynamic clusters, we
integrate the susceptibility $[s^2(\epsilon) \tanh(\epsilon/T)]/\epsilon$ of a single
cluster over the DOS (\ref{eq:DOS}). For $T < \epsilon_{\rm cross}$, $\chi_{\rm dy}$ is
dominated by the clusters with lowest energies and sizes close to $s_c$, giving
$\chi_{\rm dy} \sim {n_{s_c}}/[{T\ln^{\phi-1}(h_x/T)} ]$. For temperatures above
$\epsilon_{\rm cross}$, damping becomes unimportant and $\chi_{\rm dy}$ takes the quantum
Griffiths form $\chi_{\rm dy} \sim T^{c/b-1}$.


Finally, we consider the heat capacity which is dominated by finite size clusters, the
infinite cluster does not contribute for $T \ll J$. The heat capacity of a single dynamic
cluster with gap $\epsilon$ is \emph{not} equal to that of a two-level system, but given
by $C \approx \pi \alpha s(\epsilon) T /(3 \epsilon)$ for $T \ll \epsilon$ and $C \sim
(\epsilon/T)^{2-2s(\epsilon)\alpha}$ for $T \gg \epsilon$.\cite{CostiZarand99}
Integrating over the DOS (\ref{eq:DOS_asymp}) gives $C_{\rm dy} \sim
n_{s_c}/\ln^{\phi-1}(h_x/T)$. The frozen clusters contribute a term of the same
functional form, $C_{\rm st} \sim n_{s_c}/\ln^{\phi-1}(h_x/T)$. For $T>\epsilon_{\rm
cross}$, the logarithmic temperature dependence is replaced by the quantum Griffiths
behavior $C \sim T^{c/b}$.


To summarize, we have have shown that the diluted quantum Ising model with ohmic
dissipation undergoes a novel percolation QPT. Observables dominated by the frozen
clusters, such as total magnetization and susceptibility, show classical percolation
critical behavior. For these quantities our analysis is asymptotically exact. Observables
dominated by the fluctuating clusters, such as the heat capacity, are \emph{noncritical}
across the percolation threshold but display singular temperature dependencies (that can
be expressed in terms of the low-energy properties  of the ohmic spin-boson model).
In the remaining paragraphs we put our results into broader perspective.

In contrast to the generic transition for $p<p_c$ (transition (i) in  Fig.\
\ref{cap:Phase-diagram}b), our percolation transition (transition (ii) in  Fig.\
\ref{cap:Phase-diagram}b) is not smeared by the mechanism of Ref.\ \onlinecite{Vojta03a}
because different percolation clusters are not coupled. Instead, the frozen clusters make
an \emph{incoherent} contribution to the magnetization. Deviations from a pure
percolation scenario can change this. If tails in the interaction couple (even very
weakly) different frozen clusters, they align coherently, and the smeared transition
scenario is restored. However, for a rapidly decaying interaction, the smearing of the
percolation transition becomes important only at very low energies.

In addition to the experimental examples mentioned after (\ref{eq:Hamiltonian}), our
dissipative quantum Ising model is related to itinerant Ising magnets where the damping
is due to the electrons. Integrating out the bath modes in (\ref{eq:Hamiltonian}) leads
to a model in the same universality class as Hertz'\cite{Hertz76} field theory for the
itinerant antiferromagnetic transition.\cite{WVTC05,CugliandoloLozanoLozza05} Thus, local
ohmic dissipation correctly captures the \emph{leading} effects of the damping of the
magnetic modes in itinerant magnets. It should be emphasized, however, that itinerant
magnets contain extra complications such as long-range RKKY interactions, making a pure
percolation scenario very unlikely. In Ref.\ \onlinecite{CastroNetoJones00}, a theory
similar to our description of the classical superparamagnetic phase was suggested as an
explanation of the unusual thermodynamics\cite{Stewart01} of heavy fermion compounds.
Such a theory can be expected to hold above the smeared phase transition.\cite{Vojta03a}
Moreover, the importance of the dissipation in these systems is currently
controversial.\cite{CastroNetoJones05,MillisMorrSchmalian05,CastroNetoJones05b}

Other possible applications include two-level atoms in diluted optical lattices coupled
to an electromagnetic field, random arrays of tunneling defects in solids or, in the
future, many coupled qubits in a noisy environment.

We acknowledge discussions with H.\ Rieger, J.\ Schmalian, and M.\ Vojta.  This work has
been supported by the NSF under grant no. DMR-0339147, and by Research Corporation.

\bibliographystyle{apsrev}
\bibliography{../00bibtex/rareregions}

\begin{thebibliography}{36}
\expandafter\ifx\csname natexlab\endcsname\relax\def\natexlab#1{#1}\fi
\expandafter\ifx\csname bibnamefont\endcsname\relax
  \def\bibnamefont#1{#1}\fi
\expandafter\ifx\csname bibfnamefont\endcsname\relax
  \def\bibfnamefont#1{#1}\fi
\expandafter\ifx\csname citenamefont\endcsname\relax
  \def\citenamefont#1{#1}\fi
\expandafter\ifx\csname url\endcsname\relax
  \def\url#1{\texttt{#1}}\fi
\expandafter\ifx\csname urlprefix\endcsname\relax\def\urlprefix{URL }\fi
\providecommand{\bibinfo}[2]{#2}
\providecommand{\eprint}[2][]{\url{#2}}

\bibitem[{\citenamefont{Fisher}(1992)}]{Fisher92}
\bibinfo{author}{\bibfnamefont{D.~S.} \bibnamefont{Fisher}},
  \bibinfo{journal}{Phys. Rev. Lett.} \textbf{\bibinfo{volume}{69}},
  \bibinfo{pages}{534} (\bibinfo{year}{1992}).

\bibitem[{\citenamefont{Fisher}(1995)}]{Fisher95}
\bibinfo{author}{\bibfnamefont{D.~S.} \bibnamefont{Fisher}},
  \bibinfo{journal}{Phys. Rev. B} \textbf{\bibinfo{volume}{51}},
  \bibinfo{pages}{6411} (\bibinfo{year}{1995}).

\bibitem[{\citenamefont{Thill and Huse}(1995)}]{ThillHuse95}
\bibinfo{author}{\bibfnamefont{M.}~\bibnamefont{Thill}} \bibnamefont{and}
  \bibinfo{author}{\bibfnamefont{D.~A.} \bibnamefont{Huse}},
  \bibinfo{journal}{Physica A} \textbf{\bibinfo{volume}{214}},
  \bibinfo{pages}{321} (\bibinfo{year}{1995}).

\bibitem[{\citenamefont{Young and Rieger}(1996)}]{YoungRieger96}
\bibinfo{author}{\bibfnamefont{A.~P.} \bibnamefont{Young}} \bibnamefont{and}
  \bibinfo{author}{\bibfnamefont{H.}~\bibnamefont{Rieger}},
  \bibinfo{journal}{Phys. Rev. B} \textbf{\bibinfo{volume}{53}},
  \bibinfo{pages}{8486} (\bibinfo{year}{1996}).

\bibitem[{\citenamefont{Vojta}(2006{\natexlab{a}})}]{Vojta06}
\bibinfo{author}{\bibfnamefont{T.}~\bibnamefont{Vojta}}, \bibinfo{journal}{J.
  Phys. A} \textbf{\bibinfo{volume}{39}}, \bibinfo{pages}{R143}
  (\bibinfo{year}{2006}{\natexlab{a}}).

\bibitem[{\citenamefont{Leggett et~al.}(1987)\citenamefont{Leggett,
  Chakravarty, Dorsey, Fisher, Garg, and Zwerger}}]{LCDFGZ87}
\bibinfo{author}{\bibfnamefont{A.~J.} \bibnamefont{Leggett}},
  \bibinfo{author}{\bibfnamefont{S.}~\bibnamefont{Chakravarty}},
  \bibinfo{author}{\bibfnamefont{A.~T.} \bibnamefont{Dorsey}},
  \bibinfo{author}{\bibfnamefont{M.~P.~A.} \bibnamefont{Fisher}},
  \bibinfo{author}{\bibfnamefont{A.}~\bibnamefont{Garg}}, \bibnamefont{and}
  \bibinfo{author}{\bibfnamefont{W.}~\bibnamefont{Zwerger}},
  \bibinfo{journal}{Rev. Mod. Phys.} \textbf{\bibinfo{volume}{59}},
  \bibinfo{pages}{1} (\bibinfo{year}{1987}).

\bibitem[{\citenamefont{Weiss}(1993)}]{Weiss_book93}
\bibinfo{author}{\bibfnamefont{U.}~\bibnamefont{Weiss}},
  \emph{\bibinfo{title}{Quantum disspative systems}} (\bibinfo{publisher}{World
  Scientific}, \bibinfo{address}{Singapore}, \bibinfo{year}{1993}).

\bibitem[{\citenamefont{Hewson}(1993)}]{Hewson_book93}
\bibinfo{author}{\bibfnamefont{A.~C.} \bibnamefont{Hewson}},
  \emph{\bibinfo{title}{The Kondo Problem to Heavy Fermions}}
  (\bibinfo{publisher}{Cambridge University Press},
  \bibinfo{address}{Cambridge}, \bibinfo{year}{1993}).

\bibitem[{\citenamefont{Vojta}(2006{\natexlab{b}})}]{VojtaM06}
\bibinfo{author}{\bibfnamefont{M.}~\bibnamefont{Vojta}},
  \bibinfo{journal}{Phil. Mag.} \textbf{\bibinfo{volume}{86}},
  \bibinfo{pages}{1807} (\bibinfo{year}{2006}{\natexlab{b}}).

\bibitem[{\citenamefont{Harris}(1974)}]{Harris74b}
\bibinfo{author}{\bibfnamefont{A.~B.} \bibnamefont{Harris}},
  \bibinfo{journal}{J. Phys. C} \textbf{\bibinfo{volume}{7}},
  \bibinfo{pages}{3082} (\bibinfo{year}{1974}).

\bibitem[{\citenamefont{Stinchcombe}(1981)}]{Stinchcombe81}
\bibinfo{author}{\bibfnamefont{R.}~\bibnamefont{Stinchcombe}},
  \bibinfo{journal}{J. Phys. C} \textbf{\bibinfo{volume}{14}},
  \bibinfo{pages}{L263} (\bibinfo{year}{1981}).

\bibitem[{\citenamefont{dos Santos}(1982)}]{Santos82}
\bibinfo{author}{\bibfnamefont{R.~R.} \bibnamefont{dos Santos}},
  \bibinfo{journal}{J. Phys. C} \textbf{\bibinfo{volume}{15}},
  \bibinfo{pages}{3141} (\bibinfo{year}{1982}).

\bibitem[{\citenamefont{Senthil and Sachdev}(1996)}]{SenthilSachdev96}
\bibinfo{author}{\bibfnamefont{T.}~\bibnamefont{Senthil}} \bibnamefont{and}
  \bibinfo{author}{\bibfnamefont{S.}~\bibnamefont{Sachdev}},
  \bibinfo{journal}{Phys. Rev. Lett.} \textbf{\bibinfo{volume}{77}},
  \bibinfo{pages}{5292} (\bibinfo{year}{1996}).

\bibitem[{\citenamefont{Millis et~al.}(2001)\citenamefont{Millis, Morr, and
  Schmalian}}]{MillisMorrSchmalian01}
\bibinfo{author}{\bibfnamefont{A.~J.} \bibnamefont{Millis}},
  \bibinfo{author}{\bibfnamefont{D.~K.} \bibnamefont{Morr}}, \bibnamefont{and}
  \bibinfo{author}{\bibfnamefont{J.}~\bibnamefont{Schmalian}},
  \bibinfo{journal}{Phys. Rev. Lett.} \textbf{\bibinfo{volume}{87}},
  \bibinfo{pages}{167202} (\bibinfo{year}{2001}).

\bibitem[{\citenamefont{Millis et~al.}(2002)\citenamefont{Millis, Morr, and
  Schmalian}}]{MillisMorrSchmalian02}
\bibinfo{author}{\bibfnamefont{A.~J.} \bibnamefont{Millis}},
  \bibinfo{author}{\bibfnamefont{D.~K.} \bibnamefont{Morr}}, \bibnamefont{and}
  \bibinfo{author}{\bibfnamefont{J.}~\bibnamefont{Schmalian}},
  \bibinfo{journal}{Phys. Rev. B} \textbf{\bibinfo{volume}{66}},
  \bibinfo{pages}{174433} (\bibinfo{year}{2002}).

\bibitem[{\citenamefont{Castro~Neto and Jones}(2000)}]{CastroNetoJones00}
\bibinfo{author}{\bibfnamefont{A.~H.} \bibnamefont{Castro~Neto}}
  \bibnamefont{and} \bibinfo{author}{\bibfnamefont{B.~A.} \bibnamefont{Jones}},
  \bibinfo{journal}{Phys. Rev. B} \textbf{\bibinfo{volume}{62}},
  \bibinfo{pages}{14975} (\bibinfo{year}{2000}).

\bibitem[{\citenamefont{Vojta}(2003{\natexlab{a}})}]{Vojta03a}
\bibinfo{author}{\bibfnamefont{T.}~\bibnamefont{Vojta}},
  \bibinfo{journal}{Phys. Rev. Lett.} \textbf{\bibinfo{volume}{90}},
  \bibinfo{pages}{107202} (\bibinfo{year}{2003}{\natexlab{a}}).

\bibitem[{\citenamefont{Vojta}(2003{\natexlab{b}})}]{Vojta03b}
\bibinfo{author}{\bibfnamefont{T.}~\bibnamefont{Vojta}}, \bibinfo{journal}{J.
  Phys. A} \textbf{\bibinfo{volume}{36}}, \bibinfo{pages}{10921}
  (\bibinfo{year}{2003}{\natexlab{b}}).

\bibitem[{\citenamefont{Schehr and Rieger}(2006)}]{SchehrRieger06}
\bibinfo{author}{\bibfnamefont{G.}~\bibnamefont{Schehr}} \bibnamefont{and}
  \bibinfo{author}{\bibfnamefont{H.}~\bibnamefont{Rieger}},
  \bibinfo{journal}{Phys. Rev. Lett.} \textbf{\bibinfo{volume}{96}},
  \bibinfo{pages}{227201} (\bibinfo{year}{2006}).

\bibitem[{\citenamefont{Cugliandolo et~al.}(2005)\citenamefont{Cugliandolo,
  Lozano, and Lozza}}]{CugliandoloLozanoLozza05}
\bibinfo{author}{\bibfnamefont{L.~F.} \bibnamefont{Cugliandolo}},
  \bibinfo{author}{\bibfnamefont{G.~S.} \bibnamefont{Lozano}},
  \bibnamefont{and} \bibinfo{author}{\bibfnamefont{H.}~\bibnamefont{Lozza}},
  \bibinfo{journal}{Phys. Rev. B} \textbf{\bibinfo{volume}{71}},
  \bibinfo{pages}{224421} (\bibinfo{year}{2005}).

\bibitem[{\citenamefont{Wernsdorfer}(2001)}]{Wernsdorfer01}
\bibinfo{author}{\bibfnamefont{W.}~\bibnamefont{Wernsdorfer}},
  \bibinfo{journal}{Adv. Chem. Phys.} \textbf{\bibinfo{volume}{118}},
  \bibinfo{pages}{99} (\bibinfo{year}{2001}).

\bibitem[{\citenamefont{Prokofev and Stamp}(2000)}]{ProkofevStamp00}
\bibinfo{author}{\bibfnamefont{N.~V.} \bibnamefont{Prokofev}} \bibnamefont{and}
  \bibinfo{author}{\bibfnamefont{P.~C.~E.} \bibnamefont{Stamp}},
  \bibinfo{journal}{Rep. Progr. Phys.} \textbf{\bibinfo{volume}{63}},
  \bibinfo{pages}{669} (\bibinfo{year}{2000}).

\bibitem[{\citenamefont{Chiorescu et~al.}(2000)\citenamefont{Chiorescu,
  Wernsdorfer, M{\"u}ller, B{\"o}gge, and Barbara}}]{CWMBB00}
\bibinfo{author}{\bibfnamefont{I.}~\bibnamefont{Chiorescu}},
  \bibinfo{author}{\bibfnamefont{W.}~\bibnamefont{Wernsdorfer}},
  \bibinfo{author}{\bibfnamefont{A.}~\bibnamefont{M{\"u}ller}},
  \bibinfo{author}{\bibfnamefont{H.}~\bibnamefont{B{\"o}gge}},
  \bibnamefont{and} \bibinfo{author}{\bibfnamefont{B.}~\bibnamefont{Barbara}},
  \bibinfo{journal}{Phys. Rev. Lett.} \textbf{\bibinfo{volume}{84}},
  \bibinfo{pages}{3454} (\bibinfo{year}{2000}).

\bibitem[{\citenamefont{Hoyos and Vojta}()}]{HoyosVojta_unpublished}
\bibinfo{author}{\bibfnamefont{J.~A.} \bibnamefont{Hoyos}} \bibnamefont{and}
  \bibinfo{author}{\bibfnamefont{T.}~\bibnamefont{Vojta}},
  \bibinfo{note}{unpublished}.

\bibitem[{\citenamefont{Kosterlitz and Thouless}(1973)}]{KosterlitzThouless73}
\bibinfo{author}{\bibfnamefont{J.~M.} \bibnamefont{Kosterlitz}}
  \bibnamefont{and} \bibinfo{author}{\bibfnamefont{D.~J.}
  \bibnamefont{Thouless}}, \bibinfo{journal}{J. Phys. C}
  \textbf{\bibinfo{volume}{6}}, \bibinfo{pages}{1181} (\bibinfo{year}{1973}).

\bibitem[{\citenamefont{Thouless}(1969)}]{Thouless69}
\bibinfo{author}{\bibfnamefont{D.~J.} \bibnamefont{Thouless}},
  \bibinfo{journal}{Phys. Rev.} \textbf{\bibinfo{volume}{187}},
  \bibinfo{pages}{732} (\bibinfo{year}{1969}).

\bibitem[{\citenamefont{Cardy}(1981)}]{Cardy81}
\bibinfo{author}{\bibfnamefont{J.}~\bibnamefont{Cardy}}, \bibinfo{journal}{J.
  Phys. A} \textbf{\bibinfo{volume}{14}}, \bibinfo{pages}{1407}
  (\bibinfo{year}{1981}).

\bibitem[{\citenamefont{Kosterlitz}(1974)}]{Kosterlitz74}
\bibinfo{author}{\bibfnamefont{J.~M.} \bibnamefont{Kosterlitz}},
  \bibinfo{journal}{J. Phys. C} \textbf{\bibinfo{volume}{7}},
  \bibinfo{pages}{1046} (\bibinfo{year}{1974}).

\bibitem[{\citenamefont{Stauffer and Aharony}(1991)}]{StaufferAharony_book91}
\bibinfo{author}{\bibfnamefont{D.}~\bibnamefont{Stauffer}} \bibnamefont{and}
  \bibinfo{author}{\bibfnamefont{A.}~\bibnamefont{Aharony}},
  \emph{\bibinfo{title}{Introduction to Percolation Theory}}
  (\bibinfo{publisher}{CRC Press}, \bibinfo{address}{Boca Raton},
  \bibinfo{year}{1991}).

\bibitem[{\citenamefont{Costi and Zarand}(1999)}]{CostiZarand99}
\bibinfo{author}{\bibfnamefont{T.~A.} \bibnamefont{Costi}} \bibnamefont{and}
  \bibinfo{author}{\bibfnamefont{G.}~\bibnamefont{Zarand}},
  \bibinfo{journal}{Phys. Rev. B} \textbf{\bibinfo{volume}{59}},
  \bibinfo{pages}{12398} (\bibinfo{year}{1999}).

\bibitem[{\citenamefont{Hertz}(1976)}]{Hertz76}
\bibinfo{author}{\bibfnamefont{J.}~\bibnamefont{Hertz}},
  \bibinfo{journal}{Phys. Rev. B} \textbf{\bibinfo{volume}{14}},
  \bibinfo{pages}{1165} (\bibinfo{year}{1976}).

\bibitem[{\citenamefont{Werner et~al.}(2005)\citenamefont{Werner, V{\"o}lker,
  Troyer, and Chakravarty}}]{WVTC05}
\bibinfo{author}{\bibfnamefont{P.}~\bibnamefont{Werner}},
  \bibinfo{author}{\bibfnamefont{K.}~\bibnamefont{V{\"o}lker}},
  \bibinfo{author}{\bibfnamefont{M.}~\bibnamefont{Troyer}}, \bibnamefont{and}
  \bibinfo{author}{\bibfnamefont{S.}~\bibnamefont{Chakravarty}},
  \bibinfo{journal}{Phys. Rev. Lett.} \textbf{\bibinfo{volume}{94}},
  \bibinfo{pages}{047201} (\bibinfo{year}{2005}).

\bibitem[{\citenamefont{Stewart}(2001)}]{Stewart01}
\bibinfo{author}{\bibfnamefont{G.}~\bibnamefont{Stewart}},
  \bibinfo{journal}{Rev. Mod. Phys.} \textbf{\bibinfo{volume}{73}},
  \bibinfo{pages}{797} (\bibinfo{year}{2001}).

\bibitem[{\citenamefont{Castro~Neto and
  Jones}(2005{\natexlab{a}})}]{CastroNetoJones05}
\bibinfo{author}{\bibfnamefont{A.~H.} \bibnamefont{Castro~Neto}}
  \bibnamefont{and} \bibinfo{author}{\bibfnamefont{B.~A.} \bibnamefont{Jones}},
  \bibinfo{journal}{Europhys. Lett.} \textbf{\bibinfo{volume}{71}},
  \bibinfo{pages}{790} (\bibinfo{year}{2005}{\natexlab{a}}).

\bibitem[{\citenamefont{Millis et~al.}(2005)\citenamefont{Millis, Morr, and
  Schmalian}}]{MillisMorrSchmalian05}
\bibinfo{author}{\bibfnamefont{A.~J.} \bibnamefont{Millis}},
  \bibinfo{author}{\bibfnamefont{D.~K.} \bibnamefont{Morr}}, \bibnamefont{and}
  \bibinfo{author}{\bibfnamefont{J.}~\bibnamefont{Schmalian}},
  \bibinfo{journal}{Europhys. Lett.} \textbf{\bibinfo{volume}{72}},
  \bibinfo{pages}{1052} (\bibinfo{year}{2005}).

\bibitem[{\citenamefont{Castro~Neto and
  Jones}(2005{\natexlab{b}})}]{CastroNetoJones05b}
\bibinfo{author}{\bibfnamefont{A.~H.} \bibnamefont{Castro~Neto}}
  \bibnamefont{and} \bibinfo{author}{\bibfnamefont{B.~A.} \bibnamefont{Jones}},
  \bibinfo{journal}{Europhys. Lett.} \textbf{\bibinfo{volume}{72}},
  \bibinfo{pages}{1054} (\bibinfo{year}{2005}{\natexlab{b}}).

\end{thebibliography}

\end{document}